\documentclass[twocolumn]{aastex701}

\graphicspath{{./}{figure/}}

\shorttitle{NGC 3227 Blinks}
\shortauthors{Zhang et al.}

\submitjournal{ApJL}

\newcommand{\WIRO}{Department of Physics and Astronomy, University of Wyoming, Laramie, WY 82071, USA}
\newcommand{\ihep}{Key Laboratory for Particle Astrophysics, Institute of High Energy Physics, Chinese Academy of Sciences, 19B Yuquan Road, Beijing 100049, P.\,R. China}
\newcommand{\ucas}{School of Astronomy and Space Science, School of Physical Sciences, University of Chinese Academy of Sciences, 19A Yuquan Road, Beijing 100049, P.\,R. China}
\newcommand{\naoc}{National Astronomical Observatories of China, Chinese Academy of Sciences, 20A Datun Road, Beijing 100020, P.\,R. China}
\newcommand{\AndrewU}{SUPA School of Physics and Astronomy, North Haugh, St. Andrews, KY16 9SS, Scotland, UK}
\newcommand{\stsi}{Space Telescope Science Institute, 3700 San Martin Dr., Baltimore, MD 21218, USA}
\newcommand{\unlv}{Department of Physics \& Astronomy, University of Nevada, Las Vegas, 4505 S. Maryland Pkwy, Las Vegas, NV 89154-4002, USA}

\begin{document}

\title{NGC 3227 Blinks: An Extreme Changing-Look Event in a Seyfert Galaxy}

\author[sname=Zhang,gname=Kaiwen]{Kaiwen Zhang}
\affiliation{\WIRO}
\email[show]{kzhang1@uwyo.edu}

\author[sname=Brotherton,gname='Michael S.']{Michael S. Brotherton}
\affiliation{\WIRO}
\email{MBrother@uwyo.edu}

\author[sname=Du,gname=Pu]{Pu Du}
\affiliation{\ihep}
\email{dupu@ihep.ac.cn}

\author[sname=Bao,gname='Dong-Wei']{Dong-Wei Bao}
\affiliation{\ihep}
\affiliation{\ucas}
\email{sgkmndks@gmail.com}

\author[sname=Zhao,gname='Bi-Xuan']{Bi-Xuan Zhao}
\affiliation{\ihep}
\email{zhaobixuan@126.com}

\author[sname=Fang,gname='Feng-na']{Feng-na Fang}
\affiliation{\ihep}
\email{fangfengna@ihep.ac.cn}

\author[sname=Li,gname='Yan-Rong']{Yan-Rong Li}
\affiliation{\ihep}
\email{liyanrong@mail.ihep.ac.cn}

\author[sname=Hu,gname=Chen]{Chen Hu}
\affiliation{\ihep}
\email{huc@ihep.ac.cn}

\author[sname=Xiao,gname=Ming]{Ming Xiao}
\affiliation{\ihep}
\email{xiaoming@ihep.ac.cn}

\author[sname=Horne,gname=Keith]{Keith Horne}
\affiliation{\AndrewU}
\email{kdh1@st-andrews.ac.uk}

\author[sname=Kriss,gname='Gerard A.']{Gerard A. Kriss}
\affiliation{\stsi}
\email{gak@stsci.edu}

\author[sname=Proga,gname=Daniel]{Daniel Proga}
\affiliation{\unlv}
\email{daniel.proga@unlv.edu}

\author[sname=Wang,gname='Jian-Min']{Jian-Min Wang}
\affiliation{\ihep}
\affiliation{\ucas}
\affiliation{\naoc}
\email{wangjm@ihep.ac.cn}

\begin{abstract}
We report a rapid optical changing-look event in the Seyfert galaxy
NGC 3227: the optical continuum faded by $\sim$70\% and broad
H$\beta$ emission nearly disappeared within weeks, before recovering
within two months. Damped random walk simulations calibrated to the
source's normal variability show that such a decline is unlikely for
stochastic accretion fluctuations. The optical decline coincides with an X-ray spectral hardening event
detected by \textit{Swift}, indicating {additional line-of-sight absorption}. During the dim state, reverberation
mapping yields a clear 6--8~day lag between the continuum and broad
H$\beta$, placing the obscurer interior to the broad-line region.
Photoionization calculations require a fully ionized Compton-thick
column with $N_{\rm H} \approx 1.5\times10^{24}\,\mathrm{cm^{-2}}$ to
reproduce the optical continuum suppression. Together, these
observations demonstrate that compact, high-column-density obscurers
interior to the broad-line region can produce extreme optical
changing-look behavior on timescales of weeks.
\end{abstract}


\section{Introduction} \label{sec:intro}

Changing-look active galactic nuclei (CLAGNs) are AGNs that exhibit
dramatic changes in their spectral characteristics \citep{Ricci2023},
such as the appearance or disappearance of broad emission lines or
changing X-ray spectral slopes. Changing-look events provide direct
probes of changes in the inner accretion flow, the formation and
destruction of circumnuclear structures, and the gaseous environment
surrounding AGNs.

These transitions are generally attributed to two distinct physical
mechanisms \citep{Ricci2023}. In the first case, changes in the
intrinsic accretion flow onto the supermassive black hole (SMBH) lead to
the strengthening or weakening of the ionizing continuum, resulting in the
appearance or disappearance of broad emission lines
\citep{Runnoe2016,Graham2020}. In the second, known as changing
obscuration, variations in the line-of-sight column density modulate the
observed emission, producing pronounced spectral changes
\citep{Risaliti2009,Torricelli2014}. Such obscuration-driven events display X-ray spectral hardening, often
accompanied by suppressed soft X-ray emission and, in some cases,
blue-shifted broad absorption-line troughs in the ultraviolet
\citep{King2015,Kaastra2014}.

While most optical CLAGN events are attributed to intrinsic
accretion-rate changes \citep{Ricci2023,Jana2025,Temple2022}, direct
evidence that obscuration alone can drive large-amplitude optical
continuum and broad emission-line variability has been reported in only
a few cases \citep{Zeltyn2022,Oknyansky2021,car1}.

NGC 3227 is a nearby, well-studied Seyfert galaxy ($z = 0.003859$) at a
distance of 23.7~Mpc \citep{Bentz2023}, with a black hole mass of
$M_{\rm BH} \sim 1\times10^{7}\,M_{\odot}$ from reverberation mapping
\citep{Brotherton2020,Bentz2023} and dynamical modeling
\citep{Bentz2023,Bennert2026}. It is in an early-stage interaction with
its companion, NGC 3226. This interaction has resulted in substantial
internal reddening in the host galaxy of NGC 3227, as revealed by
H\,\textsc{i} radio mapping \citep{Mundell1995} and by the reddened
optical and ultraviolet spectral shape
\citep{Crenshaw2001,Mehdipour2021}. We account for this reddening in our
spectral decomposition (Section~\ref{subsec:host}).

NGC 3227 has a long history of X-ray variability and obscuration events,
with episodes of X-ray spectral hardening reported as early as 2000
using \textit{RXTE} observations \citep{Lamer2003,Markowitz2014}.
Subsequent X-ray monitoring has revealed additional hardening events
\citep{Markowitz2014,Beuchert2015,Turner2018,Wang2022}, suggesting that
such obscuration events recur. However, none of these episodes was
accompanied by high-cadence optical spectroscopic monitoring.

In November 2019, another X-ray hardening event in NGC 3227 was detected
by \textit{Swift} \citep{Mehdipour2021}, triggering intensive
multi-wavelength follow-up observations with \textit{HST},
\textit{XMM-Newton}, and \textit{NuSTAR}. During this campaign, the
nucleus exhibited a pronounced decline in flux from the UV to the X-ray
bands, with the suppression strongest at soft X-ray energies,
consistent with the {appearance of new obscuring gas} along the line of sight \citep{Mehdipour2021}. {Between the two target-of-opportunity epochs}, the obscurer column density increased from
$N_{\rm H} \sim 10^{22}$~cm$^{-2}$ in mid-November to
$\sim 10^{23}$~cm$^{-2}$ by early December, tracking the deepening X-ray suppression \citep{Mao2022}.

In this Letter, we present contemporaneous high-cadence optical
spectroscopic monitoring of NGC 3227 that covers this event. We describe
the observations and data reduction in Section~\ref{sec:obs}, the
spectral analysis in Section~\ref{sec:spec}, and our results in
Section~\ref{sec:results}, comprising the changing-look event itself, a
damped random walk null test, the reverberation analysis, and
photoionization modeling of the obscurer. We discuss the implied
obscuration geometry in Section~\ref{sec:discussion} and summarize in
Section~\ref{sec:summary}.

\section{Observations and Data Reduction} \label{sec:obs}

\subsection{WIRO spectroscopic monitoring} \label{subsec:wiro}

We monitored NGC 3227 between 2019 November {14} and 2020 May 29 with the
Wyoming Infrared Observatory (WIRO) 2.3~m telescope using the Long Slit
Spectrograph in remote-operation mode \citep{Findlay2016}, following the
procedures established for the Monitoring AGNs with H$\beta$ Asymmetry
(MAHA) campaigns \citep{Du2018}. The 900~line~mm$^{-1}$ grating provided
a dispersion of 1.49~\AA~pixel$^{-1}$ over 4000--7000~\AA.
Observations were carried out with a $5''$ wide slit fixed in the
north--south orientation. Each epoch typically consisted of three 5
minute exposures. At least one spectrophotometric standard star (usually
Feige 34, occasionally G191B2B) was observed per night, along with CuAr
lamps for wavelength calibration and dome flats from a uniformly
illuminated white card.

Data were reduced following the standard MAHA pipeline \citep{Du2018},
which employs IRAF v2.16 \citep{Tody1986, Tody1993} for the basic
reduction steps. One-dimensional spectra were
extracted with apertures of $\pm6\farcs8$ and background windows
spanning $7\farcs6$--$15\farcs1$ on both sides. Flux calibration and
extinction correction were applied using spectrophotometric standards,
but additional corrections were required to account for variable
atmospheric transparency and telescope tracking errors.

We adopted the [O\,\textsc{iii}]\,$\lambda$5007 narrow-line emission as
an internal flux calibrator, since it originates from an extended
narrow-line region that is stable on timescales of years
\citep{Peterson2013}. Variable seeing and tracking errors between epochs
introduce small shifts in the [O\,\textsc{iii}] line centroid, which
would propagate into the flux calibration. To mitigate this, all spectra
were convolved to a common spectral resolution before calibration. This
reduces the effective spectral resolution but does not affect the
broad-line measurements that are the focus of this work. Relative
spectrophotometric calibration was then performed using the algorithm of
\citet{Groningen1992}, with minor modifications \citep{Du2018}. We
measured the [O\,\textsc{iii}] fluxes over 5005--5050~\AA\ from spectra
obtained under photometric conditions. Final flux-calibrated spectra
were obtained by noise-weighted averaging of exposures from the same
night.

\subsection{Archival ZTF photometry} \label{subsec:ztf}

To supplement the continuum light curve and increase the sampling
cadence, we used archival $g$-band photometry of NGC 3227 from the
Zwicky Transient Facility \citep[ZTF;][]{ztf_doi}. The data were
extracted from a region centered on the nucleus, with only measurements
having \texttt{catflag}~$=0$ retained for quality control. All data were
drawn from object IDs 571109400017782, 572112300010962, and
1617102200002807, covering the heliocentric Julian date (HJD) range
2458800 to 2459010.

The inter-calibration between the ZTF $g$-band light curve and the WIRO
spectroscopic photometry at 5100~\AA\ was performed using
\texttt{PyCALI} \citep{pycali}. Before calibration, the ZTF observations
were rebinned by averaging all measurements within a single day, and
three binned epochs exceeding $3\sigma$ from the reconstructed light
curve were removed to improve data quality.

\section{Spectral Analysis} \label{sec:spec}

\subsection{Host decomposition and reddening correction} \label{subsec:host}

The reduced spectra of NGC 3227 were corrected for Milky Way extinction
using the G23 extinction curve of \citet{Gordon2023} with $R_V = 3.1$
and $E(B-V) = 0.02$~mag, then shifted to the rest frame. The AGN and
host galaxy contributions were separated using spectral decomposition
with the galaxy templates of \citet{Kinney1996}. The final host spectrum
was constructed as a linear combination of Sa, Sb, and bulge templates
with scaling factors of 1/10.9, 1/8.3, and 1/100, respectively. These
factors were fixed by two joint requirements: (i) the residual AGN
continuum remains non-negative in the lowest-flux state, and (ii) the
host stellar absorption features --- in particular the Na\,\textsc{i}~D
doublet near 5900~\AA\ --- are removed in the residual AGN spectra.
The host contribution is dominated by the Sa and Sb components, with a
negligible bulge contribution. This host spectrum, assumed constant
across epochs, was subtracted from all spectra.

NGC 3227 is known to suffer significant internal reddening
\citep{Crenshaw2001,Mundell1995}. Following \citet{Mehdipour2021}, we
applied the SMC Bar extinction curve of \citet{Gordon2003} with
$R_V = 4.0$ and $E(B-V) = 0.45$~mag, which they found to best
characterize the dust extinction in the nuclear region. The AGN-only
spectra were dereddened using this extinction curve to recover the
intrinsic AGN continuum and emission-line fluxes.

\subsection{Mean and rms spectra} \label{subsec:meanrms}

\begin{figure*}[ht!]
\plotone{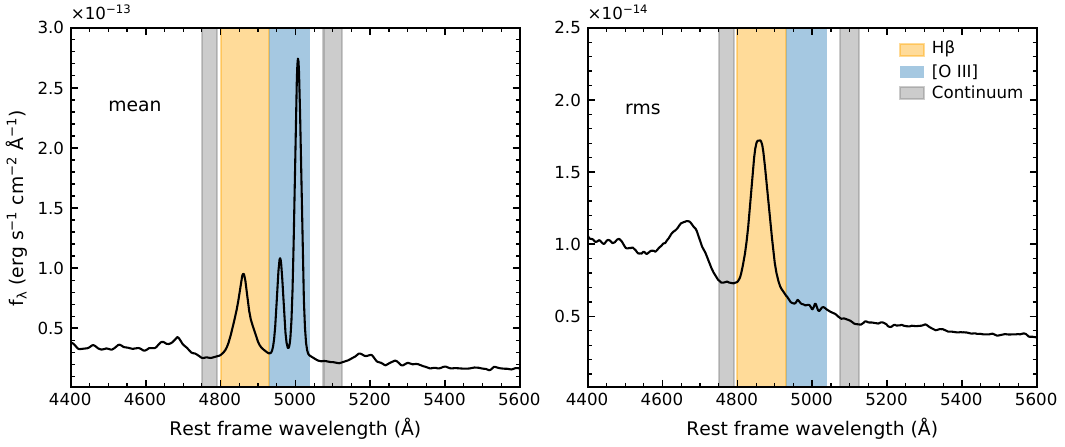}
\caption{Mean and rms spectra from the WIRO monitoring campaign of
NGC 3227. Left: mean spectrum. Right: rms spectrum. The gray region
marks the local continuum window used for H$\beta$ pseudo-continuum
subtraction, the orange region marks the H$\beta$ measurement window
(4800--4929~\AA), and the cyan region marks the [O\,\textsc{iii}]
$\lambda\lambda$4959, 5007 doublet (4929--5040~\AA). The minimal
[O\,\textsc{iii}] variability in the rms spectrum confirms the stability
of the flux calibration.
\label{fig:meanrms}}
\end{figure*}

We constructed mean and root-mean-square (rms) spectra from the
reddening-corrected AGN spectra to characterize the average emission
profile and its variability. Integration windows were then defined for
the H$\beta$ emission line and the [O\,\textsc{iii}]
$\lambda\lambda$4959, 5007 doublet, with boundaries set from the extent
of the line emission in the mean and rms spectra. The H$\beta$ profile
was measured over 4800--4929~\AA, the [O\,\textsc{iii}] doublet over
4929--5040~\AA, and the 5100~\AA\ continuum over 5075--5125~\AA. The
minimal [O\,\textsc{iii}] in the rms spectrum confirms the
stability of the flux calibration. The mean and rms spectra, together
with the wavelength windows defined above, are shown in
Figure~\ref{fig:meanrms}.

\subsection{Recovery of the broad H$\beta$ component} \label{subsec:broadhb}

\begin{figure}[ht!]
\centering
\includegraphics[width=\columnwidth]{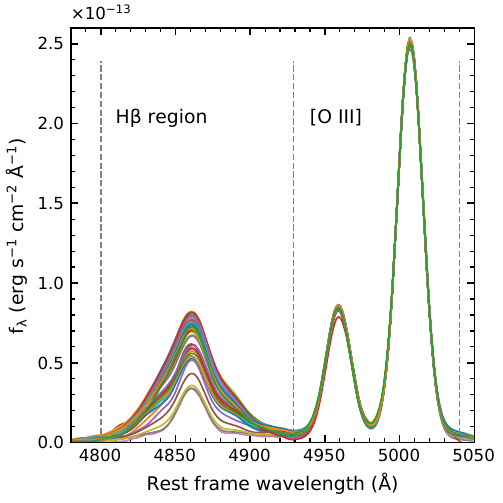}
\caption{Continuum-subtracted H$\beta$ and [O\,\textsc{iii}]
emission-line profiles, one colored curve per epoch of the 2019--2020
campaign, after subtraction of the host galaxy and a local linear
continuum (windows at 4750--4790 and 5075--5125~\AA). The H$\beta$
profile, containing both broad and narrow-line components, varies by more than a factor of two between epochs, while the
narrow [O\,\textsc{iii}] $\lambda\lambda$4959, 5007 doublet stays
essentially constant, confirming [O\,\textsc{iii}] as a stable internal
flux reference. Vertical dashed lines mark the H$\beta$ (4800--4929~\AA)
and [O\,\textsc{iii}] (4929--5040~\AA) windows.
\label{fig:nocontinuum}}
\end{figure}

To remove the local AGN continuum around the H$\beta$ line,
we defined continuum windows on either side of the
H$\beta$+[O\,\textsc{iii}] complex, specifically 4750--4790~\AA\ and
5075--5125~\AA. These windows were chosen to coincide with line-free
regions in the mean and rms spectra, as identified by visual inspection.
The local continuum was then modeled by linear interpolation between the
two windows, and subtracted to isolate the emission-line flux. The
continuum-subtracted spectra showing the broad H$\beta$ variability and
the stable [O\,\textsc{iii}] doublet across all monitoring epochs are
shown in Figure~\ref{fig:nocontinuum}.

The broad H$\beta$ component was deblended by subtracting the
narrow-line contribution from the continuum-subtracted H$\beta$ emission
region. The narrow H$\beta$ flux and profile were assumed to be stable
during the monitoring campaign, as indicated by the constancy of the
[O\,\textsc{iii}] emission line. To determine the narrow H$\beta$
profile, we used a low-state spectrum (2019 December 19) to minimize
contamination from the broad H$\beta$ component and residual continuum.
Emission-line profiles were modeled using a Markov chain Monte Carlo
(MCMC) approach to fully explore the parameter space of the broad and
narrow components. The fitting was restricted to a narrow wavelength
interval of 4840--4880~\AA. This restriction is required because the
broad H$\beta$ emission exhibits complex substructure in the lowest
state that cannot be adequately represented by a single Gaussian;
fitting the full profile would otherwise bias the decomposition by
allowing the broad component to dominate the $\chi^{2}$ minimization.
The posterior distributions for the narrow H$\beta$ component yield
amplitude $A = 2.75^{+0.21}_{-0.24}\times10^{-14}$~erg~cm$^{-2}$~s$^{-1}$~\AA$^{-1}$,
mean $\mu = 4860.62^{+0.24}_{-0.23}$~\AA, and width
$\sigma = 8.74^{+0.69}_{-0.62}$~\AA. The broad H$\beta$ profiles were
then recovered by subtracting the best-fitting narrow component from the
observed spectra. The resulting narrow-line flux ratio of
[O\,\textsc{iii}]\,$\lambda$5007/H$\beta \approx 9$ is consistent with
the typical value of $\sim$10 observed among AGNs
\citep{Kewley2006,Stern2013}, supporting the robustness of the
decomposition.

\subsection{Light curves} \label{subsec:lc}

\begin{figure*}[ht!]
\plotone{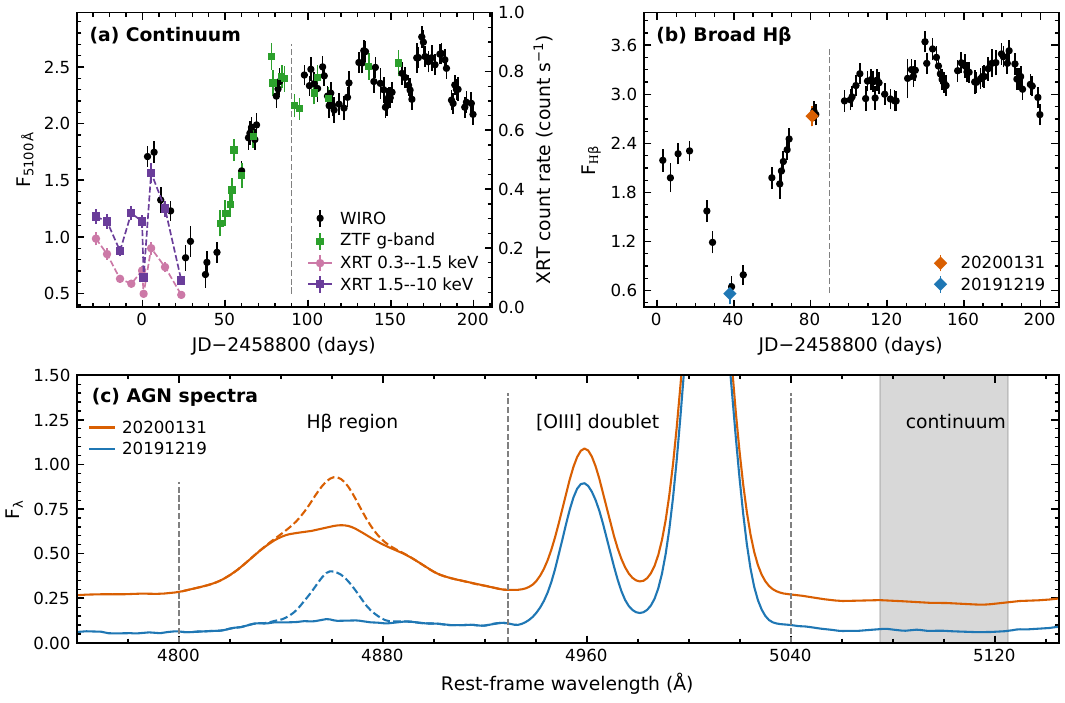}
\caption{Optical continuum, {Swift/XRT}, and broad H$\beta$ variability in NGC 3227.
(a) Light curve of the rest-frame 5100~\AA\ AGN continuum in units of
$10^{-14}$~erg~cm$^{-2}$~s$^{-1}$~\AA$^{-1}$. Black points show WIRO
measurements, while green points indicate ZTF $g$-band photometry
inter-calibrated to the WIRO 5100~\AA\ flux scale. {Pink and purple points show the \textit{Swift}/XRT 0.3--1.5 and 1.5--10~keV count rates, respectively, from \citet{Mao2022} (right-hand axis, in count~s$^{-1}$); dashed lines connect consecutive epochs.} The vertical dashed
line at 90~days marks the transition from the obscuration phase to the
subsequent normal phase.
(b) Broad H$\beta$ emission-line light curve from WIRO spectroscopy in
units of $10^{-12}$~erg~cm$^{-2}$~s$^{-1}$. Colored points highlight
epochs corresponding to the spectra shown in panel (c), using the same
color coding.
(c) Representative AGN intrinsic spectra in the low state (2019 December
19; blue) and high state (2020 January 31; orange) in units of
$10^{-14}$~erg~cm$^{-2}$~s$^{-1}$~\AA$^{-1}$. The H$\beta$ region and
the [O\,\textsc{iii}] doublet are shown. Dashed curves indicate the
narrow-line H$\beta$ emission, while solid curves represent the spectrum after H$\beta$ narrow-line subtraction. Vertical dashed lines
mark the boundaries of the H$\beta$ and [O\,\textsc{iii}] doublet
regions. The gray shaded region marks the 5100~\AA\ continuum flux at
the given epochs.
\label{fig:fig1}}
\end{figure*}

Given the complex structure of the H$\beta$ profile, the broad H$\beta$
flux in the WIRO spectra was measured by direct integration over the
recovered broad-line spectrum within the H$\beta$ window. The 5100~\AA\
continuum flux was estimated as the mean flux density within the
5075--5125~\AA\ continuum window. The total uncertainties consist of two
components: Poisson noise from the observations and systematic
uncertainties. Poisson noise was propagated throughout the data
analysis. Systematic uncertainties were estimated using two
complementary approaches, combined in quadrature, following standard
MAHA practices \citep{Du2018,Brotherton2020}. The first quantifies
within-night variations: for epochs with multiple exposures, the scatter
among individual flux measurements captures the effect of changing
atmospheric conditions and tracking errors. The second captures short-term variability across epochs: the light curve is smoothed with a five-point median filter \citep{Du2014,Du2015}, and the standard deviation of the
residuals provides an estimate of rapid intrinsic fluctuations. The
resulting broad H$\beta$ and 5100~\AA\ continuum light curves are shown
in Figure~\ref{fig:fig1}.

\section{Results} \label{sec:results}

\subsection{An extreme optical changing-look event} \label{subsec:cl}

Our optical spectroscopic monitoring with WIRO, which overlaps with the
2019 November multi-wavelength campaign, shows a contemporaneous decline
in optical flux that evolves on weekly timescales
(Figure~\ref{fig:fig1}). The AGN optical continuum at 5100~\AA\ and the
broad H$\beta$ light curve show substantial declines of approximately
70\% and 80\%, respectively. The broad H$\beta$ profile nearly
disappears in the low state (2019 December 19) and recovers by the
high state (2020 January 31), while the narrow H$\beta$ emission remains
unchanged throughout. \textbf The {\textit{Swift}/XRT hardening began before the first WIRO epoch and continued
through the early optical drop. During the period of joint coverage, 
both the optical continuum and the \textit{Swift}/XRT count rates declined, 
though the X-ray sampling is sparse and ends prior to the optical minimum. The overlapping X-ray hardening supports a physical association with the optical decline, but we cannot conclusively demonstrate that both behaviors were produced by the exact same obscuring structure.}

Given this rapid decline in continuum flux, stochastic accretion
variability alone is an unlikely explanation: only $\sim$0.2\% of damped
random walk simulations calibrated to NGC 3227's normal-state
variability reproduce a decline this steep (Section~\ref{subsec:drw}).
Combined with the X-ray hardening at the onset of the optical decline,
we argue that this event is driven by obscuration. Reverberation mapping through the dim state
(Section~\ref{subsec:lags}) recovers a robust 6--8~day lag between
the continuum and broad H$\beta$, showing that the broad-line region
continues to reverberate and placing the obscuring material interior to
the broad-line region.

\subsection{Reverberation lags} \label{subsec:lags}

We divided the light curves into two temporal segments. The obscuration
phase (2019 November {14} to 2020 January 31) captures the sharp decline
in H$\beta$ flux, following an X-ray hardening event, and its subsequent
recovery. The normal phase (2020 February 2 to 2020 May 29) reflects
typical AGN stochastic variability.

We employed the interpolated cross-correlation function
\citep[ICCF;][]{Gaskell}, implemented with the \texttt{PyCCF} package
\citep{pyccf}, to estimate the time lag between the ionizing continuum
and the H$\beta$ emission line, thereby inferring the broad-line region
(BLR) size via the light-travel distance. To complement this classical
approach, we also applied Multiple and Inhomogeneous Component Analysis
\citep[MICA;][]{mica}, a Bayesian reverberation-mapping framework. MICA
simultaneously models the stochastic continuum and the line response
using a flexible transfer-function parameterization.

To improve coverage during the obscuration phase, we augmented the
5100~\AA\ continuum with inter-calibrated ZTF data, as sparse sampling
degrades both ICCF and MICA results more than the modest systematics
introduced by inter-calibration. In contrast, ZTF data were
excluded from the normal phase to preserve the homogeneity and precision
of our primary dataset, though including the inter-calibrated ZTF data in
the normal phase yields an essentially identical lag. ICCF and MICA
analyses were performed independently for each phase, enabling a direct
comparison of the delay behavior.

\begin{figure*}[ht!]
\plotone{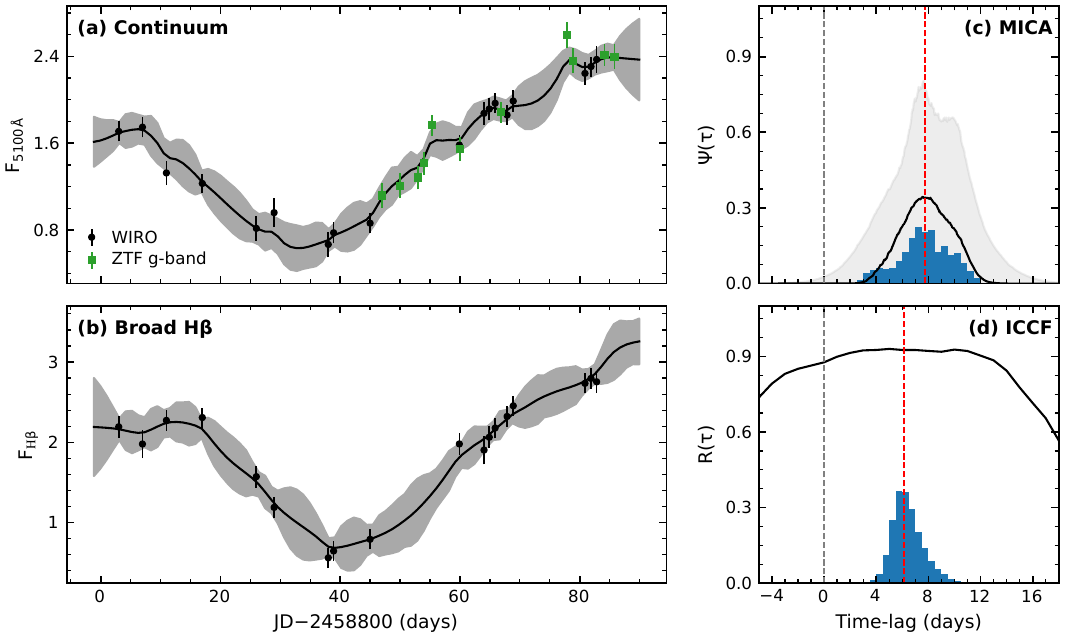}
\caption{Time-lag analysis of the continuum and broad H$\beta$ emission
during the obscuration phase. Panels (a) and (b) show the observed
(points) and MICA-reconstructed light curves of the rest-frame
5100~\AA\ continuum and the broad H$\beta$ emission, in units of
$10^{-14}$~erg~cm$^{-2}$~s$^{-1}$~\AA$^{-1}$ and
$10^{-12}$~erg~cm$^{-2}$~s$^{-1}$, respectively. Solid black curves
trace the median reconstructed light curve and gray bands show its
$1\sigma$ uncertainty. The 5100~\AA\ continuum in panel (a) is
supplemented with the inter-calibrated ZTF $g$-band photometry (green
points). Panel (c) shows the MICA lag posterior distribution
(histogram); the solid black curve is the median reconstructed transfer
function and the gray band marks its $1\sigma$ uncertainty. Panel (d)
shows the ICCF centroid distribution (histogram) together with the
cross-correlation function (solid black curve). Vertical red dashed
lines mark the median lag values. The MICA lag posterior distribution
has a median of $\sim$7.7~d, while the ICCF centroid distribution has a
median value of $\sim$6~d.
\label{fig:CCF2}}
\end{figure*}

\begin{figure*}[ht!]
\plotone{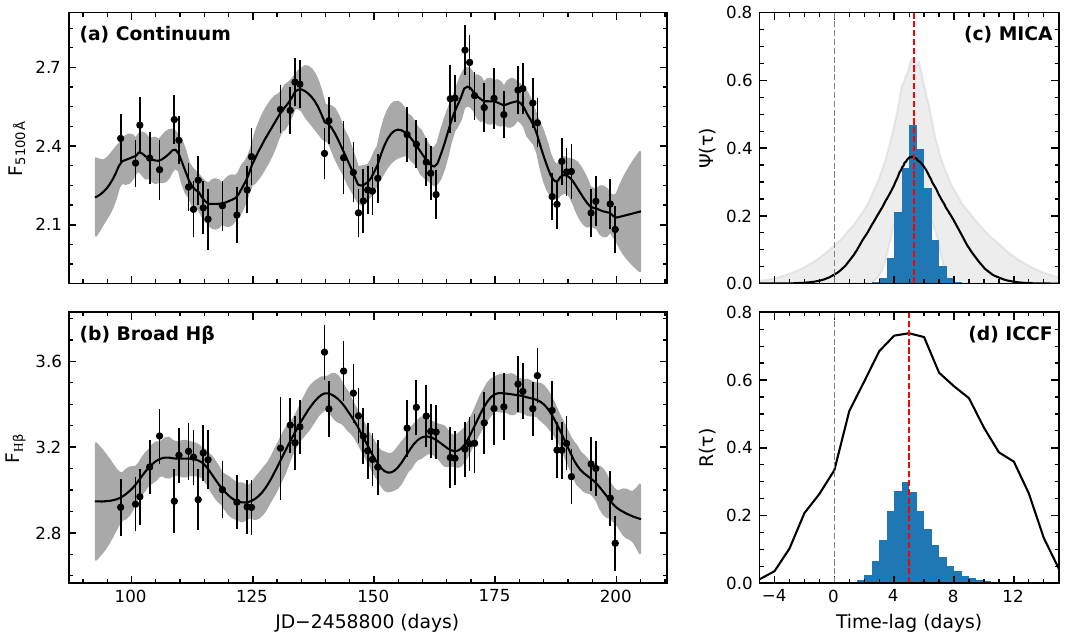}
\caption{Time-lag analysis of the continuum and broad H$\beta$ emission
during the normal phase. Same as Figure~\ref{fig:CCF2}, but for the
normal phase. Both methods yield a consistent reverberation lag of
$\sim$5~d, with the MICA lag posterior distribution having a median of
$\sim$5.4~d, indicating a stable broad-line region response under
presumably unobscured conditions.
\label{fig:CCF1}}
\end{figure*}

During the obscuration phase (Figure~\ref{fig:CCF2}), the ICCF centroid
distribution gives a lag of {$6.13^{+1.44}_{-0.99}$~days}, while MICA
yields a broader posterior at $7.72^{+2.09}_{-2.10}$~days.
During the normal phase (Figure~\ref{fig:CCF1}), ICCF analysis yields a
lag of {$4.99^{+1.53}_{-1.43}$~days} and MICA gives
$5.38^{+0.87}_{-0.88}$~days; both methods converge on a consistent lag
of $\sim$5~days, indicating a relatively stable BLR response under unobscured conditions and consistent with previous reverberation mapping results for NGC 3227 \citep{Bentz2023,Brotherton2020}.

\subsubsection{Robustness of the obscuration-phase lag}

To assess whether the deep-valley shape and reduced cadence of the
obscuration phase could bias the recovered lag, we performed mock
recovery tests using ICCF. We treated the MICA-reconstructed mean
continuum light curve as the deterministic ``true'' continuum, convolved
it with a Gaussian transfer function of standard deviation
$\sigma \approx 1.67$~days to generate an H$\beta$ light curve, interpolated
both at the actual observation epochs, and assigned per-epoch
uncertainties by rescaling the observed flux errors so that each mock
point preserves the fractional uncertainty of the corresponding real
measurement. We tested two input lags: $\tau_0 = 0$ (null) and
$\tau_0 = 6$~days (recovery). ICCF recovered both inputs within their
quoted uncertainties ($\tau = 0.11^{+0.94}_{-0.53}$~d for $\tau_0 = 0$
and $\tau = 5.50^{+0.98}_{-0.97}$~d for $\tau_0 = 6$~d), demonstrating
that the obscuration-phase lag is not a spurious artifact of the
sampling. Drawing pairs from the observed and mock zero-lag
cross-correlation centroid distributions, we find
$p(\tau_{\rm null}\ge\tau_{\rm obs})\approx 6.8\times10^{-5}$,
equivalent to a $3.8\sigma$ rejection of the null hypothesis
$\tau = 0$; the observed lag distribution is therefore inconsistent with
arising from a zero-lag light curve under the actual sampling and noise.

\subsection{Damped random walk null test} \label{subsec:drw}

To assess whether the observed $\sim$70\% optical continuum decline
could arise from stochastic accretion fluctuations rather than an
external obscuration event, we performed a null-hypothesis test using
damped random walk (DRW) simulations.

We used five seasons of MAHA optical monitoring of NGC 3227 obtained
outside the obscuration episode: the 2017 season \citep{Brotherton2020}
and four subsequent seasons spanning 2019--2023 ({Zhang et al.\ 2027, MAHA VI, in preparation}). A monotonically declining segment of
the 2023 season was excluded to ensure that the DRW parameters reflect
stochastic variability. For each season, we independently inferred the
DRW parameters, $\sigma$ (variability amplitude) and $\tau$
(characteristic timescale), and combined the posterior samples to
determine representative median values. We obtained
$\tau = 9.66^{+8.89}_{-3.69}$~days and
$\sigma = 3.63^{+1.87}_{-1.67}\times10^{-15}$~erg~cm$^{-2}$~s$^{-1}$~\AA$^{-1}$.

Using these median parameters, we generated 100,000 mock light curves
drawn from a DRW Gaussian process, adopting the same temporal sampling
and photometric uncertainties as during the monitoring campaign. For
each realization, we searched for events that exhibited a $\geq$70\% decline in continuum flux on timescales shorter than 40 days. Such events occur in only 0.2\% of the simulated light curves, and
only a similarly small fraction of light curves would recover to a high
state on the same timescale. This low occurrence rate indicates that the
observed continuum drop and subsequent rise is very unlikely to arise from stationary stochastic accretion variability alone.

\subsection{Photoionization modeling of the obscurer} \label{subsec:cloudy}

The host-galaxy subtracted rest-frame 5100~\AA\ continuum flux drops from
$2.24\times10^{-14}$ to
$6.68\times10^{-15}$~erg~cm$^{-2}$~s$^{-1}$~\AA$^{-1}$ between the high
and low states, a decline of $\sim$70\%. To interpret this attenuation,
we performed photoionization modeling with \texttt{Cloudy} C25
\citep{cloudy} to identify the obscurer column density required to
reproduce the observed low-state 5100~\AA\ flux. The ionizing source was modeled
with the intrinsic AGN spectral energy distribution of NGC 5548
\citep{ngc5548}, normalized to the high-state luminosity
$\lambda L_{5100\,\text{\AA}} \simeq 7.76\times10^{42}$~erg~s$^{-1}$.
The obscurer was modeled as a plane-parallel slab of solar-abundance
gas.

The exact location and density of the obscurer are not directly
constrained by our data; therefore, we choose sampling ranges that are
physically plausible for an obscurer interior to the 5100~\AA\
continuum-emitting region. Following the disk size--luminosity relation
of \citet{Burke2021}, the characteristic 2500~\AA\ emitting radius of
NGC 3227 is $\sim$0.09~light-days, and the empirical continuum
reverberation scaling $\tau \propto \lambda^{4/3}$
\citep{Edelson2015,Fausnaugh2016} extrapolates this to
$\sim$0.23~light-days at 5100~\AA. We therefore sampled obscurer inner
radii of $r = 0.1$--0.2~light-days, just interior to the 5100~\AA\
emitting region, and hydrogen number densities
$\log (n_{\rm H}/\mathrm{cm^{-3}}) = 9$--10, typical of gas at BLR scales
\citep{Peterson2006}.

We performed grid photoionization modeling over this parameter space,
with column densities spanning
$\log (N_{\rm H}/\mathrm{cm^{-2}}) = 23$--24.5.
Across this grid the ionization parameter spans
$\log U \approx 3$--$4.5$, leaving the obscurer fully ionized in all
cases. The transmitted 5100~\AA\ flux therefore depends only on
$N_{\rm H}$, with negligible sensitivity to $n_{\rm H}$ or $r$.
The \texttt{Cloudy} results coincide with the
analytic Thomson prediction
$T = \exp(-1.17\,N_{\rm H}\,\sigma_{\rm T})$ for fully ionized
hydrogen\,+\,helium gas at solar abundance, confirming that bound-free
and line opacities are negligible at optical wavelengths and that
electron scattering accounts for the entire attenuation. Reproducing the
observed flux decrease requires a Compton-thick column density of
$N_{\rm H} \simeq 1.5\times10^{24}$~cm$^{-2}$. Across the explored
\texttt{Cloudy} grid, energy is conserved to within $\sim$3\%; at the
best-fit column the obscurer's diffuse outward continuum at 5100~\AA\
remains below $\sim4\times10^{-4}$ of the transmitted flux, confirming
that the fully ionized cloud does not self-illuminate the optical band
probed in this work.

\section{Discussion} \label{sec:discussion}

The detection of a coherent reverberation lag between the continuum and
broad H$\beta$ during the obscuration phase localizes the obscuring
material: the H$\beta$ decline is the BLR's light-travel response to an
attenuated ionizing continuum, not a direct extinction of the
line-emitting gas along our sightline, so the BLR continues to reprocess
the continuum it receives. This excludes obscuration exterior to the
BLR. Such an obscurer dims H$\beta$ only by attenuating the line
emission, so suppressing the broad H$\beta$ flux by $\sim$80\% requires
it to cover most of the spatially extended BLR ---
a structure comparable in scale to the BLR itself. If the obscurer covers the continuum source and BLR simultaneously (for
example, an in-situ rise in opacity along the sightline), the two
decline together, producing correlated, zero-lag drops. 

If instead the obscurer transits the
system, it occults the continuum source and the surrounding BLR at
different times and smears the H$\beta$ decline over the crossing --- in
neither case reproducing the coherent $\sim$6--8~day delay we observe.
A well-defined positive lag, in which the entire H$\beta$ light curve
follows the continuum after a fixed delay, is instead the signature of
light-travel reprocessing and requires the obscurer to lie interior to
the BLR, where it attenuates the continuum before it reaches the
line-emitting gas. This inner location is consistent with the X-ray
analysis of the 2019 event by \citet{Mao2022}, and with this system's
earlier X-ray occultations, whose absorbing clouds were also placed
within or at the inner edge of the BLR \citep{Turner2018,Beuchert2015}.

Photoionization modeling further shows that reproducing the observed
attenuation of the 5100~\AA\ continuum requires a fully ionized
Compton-thick column density of
$N_{\rm H} \sim 1.5\times10^{24}$~cm$^{-2}$ along the line of sight.
Such a large column density is qualitatively consistent with the
pronounced soft X-ray suppression and X-ray spectral hardening observed
at the onset of this event, although the X-ray-derived column density is
one to two orders of magnitude lower. This difference likely reflects
both temporal and geometric factors: {the X-ray column was likely still
\textit{evolving} when last measured}, increasing from $\sim10^{22}$ to
$\sim10^{23}$~cm$^{-2}$ over the weeks before the optical minimum
\citep{Mao2022}, and no
X-ray observation samples the epoch of deepest optical suppression
(2019 December 19, both later and deeper than the final X-ray epoch),
precluding a direct comparison at the same evolutionary stage.
Furthermore, the X-ray column density traces a narrow sightline to the
compact corona, whereas the optical continuum sees a sightline to the
continuum source at a larger radius, where the obscurer's column density may differ.

\begin{figure*}[ht!]
\plotone{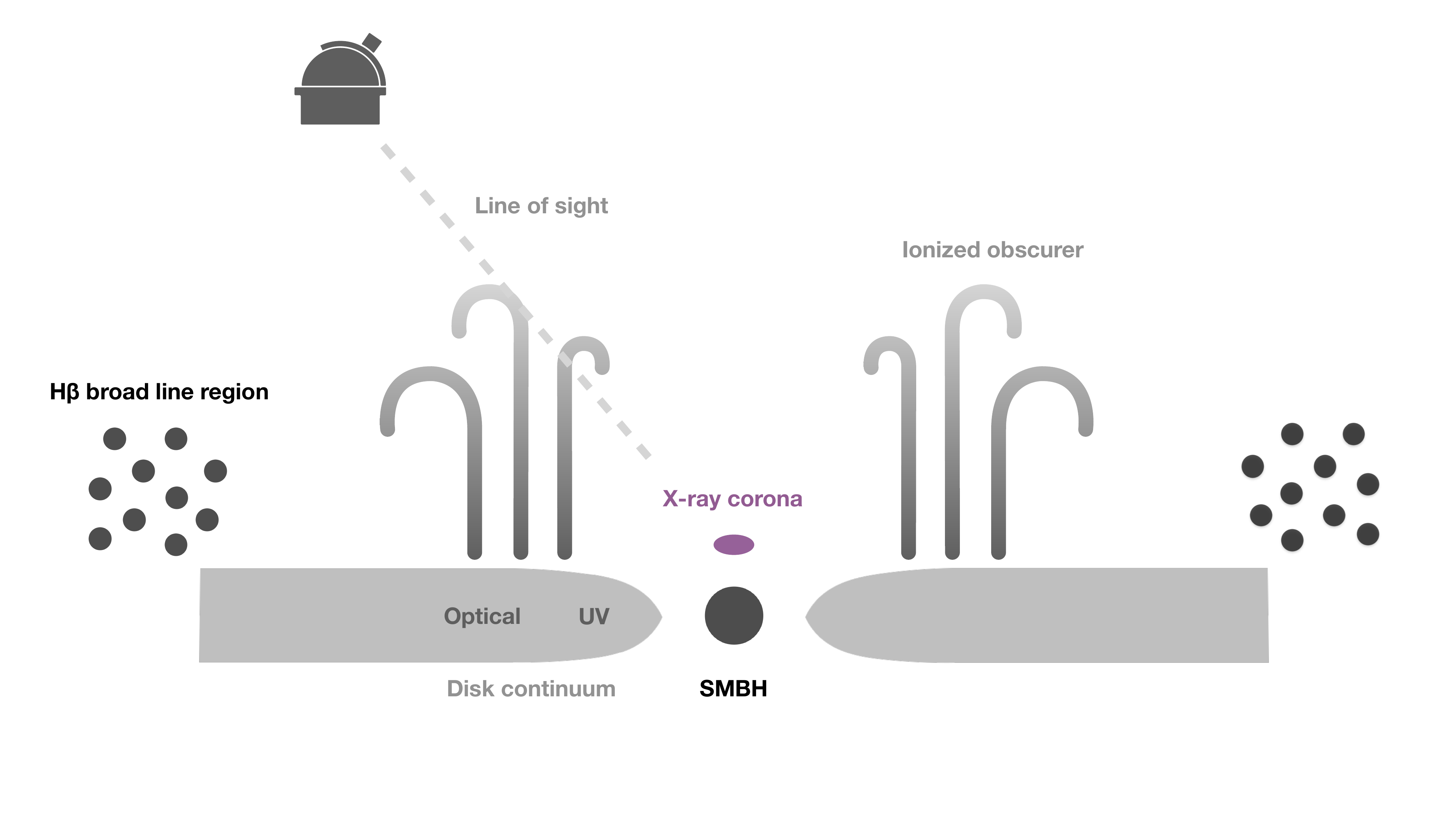}
\caption{Schematic illustration of the proposed obscuration geometry in
NGC 3227. A dense, Compton-thick obscuring structure located interior to
the BLR rises above the accretion disk, blocking the ionizing continuum
before it reaches the BLR while also attenuating the observed 5100~\AA\
optical continuum along our line of sight.
\label{fig:cartoon}}
\end{figure*}

To illustrate the observational results, we present a proposed
obscuration geometry in Figure~\ref{fig:cartoon}, which is similar to
that previously offered for Mrk 817 \citep{kara}, where a comparable
inner obscurer left the broad UV lines decoupled from the observed
continuum (a ``holiday''; \citealt{Homayouni}). In NGC 3227 the
covering instead suppresses the continuum reaching the
H$\beta$-emitting BLR, producing a full H$\beta$ changing-look event.
Unlike the Mrk 817 holiday, however, the persistent $\sim$6--8~day lag
shows that H$\beta$ still reverberates in response to the attenuated
continuum. The physical nature of the obscurer ---
whether a disk wind or another circumnuclear structure --- cannot be
determined from the present data and remains an open question. We note,
however, that a time-dependent, line-driven accretion-disk wind is one
plausible candidate: simulations show that such winds can become
Compton-thick along equatorial sightlines \citep{proga,proga2004}.

{The event in NGC 3227 is currently unique in exhibiting X-ray
obscuration accompanied by an optical changing-look transition, in which
the broad emission lines nearly vanished alongside the continuum. In
contrast, most X-ray changing-obscuration events do not show a clear
correlation between the X-ray and UV/optical changing-look behavior.
Certain X-ray changing-obscuration events (e.g., Mrk 817 and NGC 5548)
instead exhibit (a) blueshifted absorption-line troughs rather than the
near-complete continuum and emission-line suppression observed here, and
(b) a BLR ``holiday,'' in which the broad lines decouple from the
ionizing-continuum variability \citep{goad2016,dehghanian2019,kara,Homayouni}. Such holidays may depend on the obscurer's column density ($N_{\rm H}$): if
the material is transparent to specific bands of the spectral energy
distribution, only partial suppression occurs. Because Compton-thick,
near-total-coverage obscuration events are likely far rarer than
ordinary obscuration, NGC 3227 remains the only unambiguous case
identified to date.}

\section{Summary} \label{sec:summary}

We have presented high-cadence optical spectroscopic monitoring of an
extreme changing-look event in NGC 3227 and shown that: (1) the
5100\,\AA\ continuum and broad H$\beta$ flux fell by $\sim$70\% and
$\sim$80\% and recovered within two months, a decline that damped random
walk simulations disfavor as stochastic accretion variability
($p=0.002$) and that coincides with \textit{Swift} X-ray hardening;
(2) a persistent 6--8\,day continuum-to-H$\beta$ lag through the dim
state shows the broad-line region still reverberates, placing the
obscurer interior to the BLR; and (3) reproducing the optical
suppression requires a fully ionized, Compton-thick column,
$N_{\rm H}\simeq1.5\times10^{24}$\,cm$^{-2}$. Together these establish
that compact, high-column-density obscuration interior to the BLR can
drive an extreme optical changing-look event on week timescales. Such
events are hard to characterize from either band alone --- X-ray and UV
data fix the absorber's column and ionization state, high-cadence
optical spectroscopy localizes it through reverberation mapping --- so
simultaneous coverage is needed to constrain the obscurer's geometry.
Wide-field time-domain surveys with the Rubin Observatory should uncover
many more and motivate the coordinated follow-up such events require.

\begin{acknowledgments}
We thank the WIRO engineering staff James Weger and Conrad Vogel for
their dedicated technical support in maintaining the telescope and
instrumentation that made these observations possible. Based on
observations obtained with the Samuel Oschin Telescope 48 inch and the
60 inch Telescope at the Palomar Observatory as part of the Zwicky
Transient Facility project. ZTF is supported by the National Science
Foundation under Grants No. AST-1440341 and AST-2034437 and a
collaboration including current partners Caltech, IPAC, the Weizmann
Institute for Science, the Oskar Klein Center at Stockholm University,
the University of Maryland, Deutsches Elektronen-Synchrotron and
Humboldt University, the TANGO Consortium of Taiwan, the University of
Wisconsin at Milwaukee, Trinity College Dublin, Lawrence Livermore
National Laboratories, IN2P3, University of Warwick, Ruhr University
Bochum, Northwestern University and former partners the University of
Washington, Los Alamos National Laboratories, and Lawrence Berkeley
National Laboratories. Operations are conducted by COO, IPAC, and UW.

This work was supported by the U.S. National Science Foundation through
grant AST-2511198. P.D. acknowledges financial support from the National
Key R\&D Program of China (2023YFA1607903 and 2021YFA1600404), from NSFC
grants NSFC-12521005, -12022301 and -11991051, the National Science and
Technology Major Project (2024ZD0300303), and by the China Manned Space
Project with no. CMS-CSST-2025-A07.
\end{acknowledgments}

\facilities{WIRO (Wyoming Infrared Observatory), ZTF (Zwicky Transient Facility), Swift}

\software{Cloudy \citep{cloudy},
          IRAF \citep{Tody1986,Tody1993},
          MICA \citep{mica},
          PyCALI \citep{pycali},
          PyCCF \citep{pyccf}
          }

\section*{Data Availability}
The underlying data, light curves, spectra, and processing/analysis scripts used to produce the figures and results presented in this manuscript are available via \dataset[DOI: 10.5281/zenodo.23191065]{https://doi.org/10.5281/zenodo.23191065}.

\clearpage
\bibliography{new.ms.bib}{}
\bibliographystyle{aasjournalv7}

\end{document}